\documentclass[preprint,showpacs]{revtex4}
\usepackage{amssymb}
\usepackage{graphicx}
\usepackage{bm}
\pdfoutput=1
\usepackage{epstopdf}
\begin{document}

\title{The Contact in the BCS-BEC crossover for finite range interacting ultracold Fermi gases}
\author{Santiago F. Caballero-Ben\'{\i}tez, Rosario Paredes and V\'{\i}ctor Romero-Roch\'{\i}n\footnote{Corresponding author: romero@fisica.unam.mx} }

\affiliation{ 
 Instituto de F\'{\i}sica, Universidad
Nacional Aut\'onoma de M\'exico, Apartado Postal 20-364, M\'exico D.
F. 01000, M\'exico. }


\date{\today}
\begin{abstract}

Using mean-field theory for the Bardeen-Cooper-Schriefer (BCS) to the Bose-Einstein condensate (BEC) crossover we investigate the ground state thermodynamic properties of an interacting homogeneous Fermi gas. The interatomic interactions modeled through a finite range potential allows us to explore the entire region from weak to strong interacting regimes with no approximations. To exhibit the thermodynamic behavior as a function of the potential parameters in the whole crossover region, we concentrate in studying the Contact variable, the thermodynamic conjugate of the inverse of the $s$-wave scattering length. Our analysis allows us to validate the mean-field approach across the whole crossover.
It also leads to predict a quantum phase transition-like in the case when the potential range becomes large. This finding is a direct consequence of the $k$-dependent energy gap for finite interaction range potentials.

\end{abstract}
\maketitle

Experiments performed on dilute ultracold two-component Fermi gases represent perhaps the cleanest scenario that allows to access the crossover from Bardeen-Cooper-Schrieffer (BCS) superfluidity of Cooper pairs to Bose-Einstein condensation (BEC) of tightly bound fermion pairs\cite{Greiner,Jochim,Zwierlein,Bourdel,Kinast,Partridge,Stewart}. These systems can be modeled as a gas of interacting fermions in a homogeneous environment. Inhomogeneities created by a confining potential may be included via the local density approximation\cite{Dalfovo,Bulgac1}. 
Although the study of the crossover has been extensively addressed in the literature at zero \cite{Heiselberg,Carlson,Chang,Carlson2,Bruun,Perali,Pieri,Astrakharchick,Hu,Nishida,Strecher,Jauregui,Arnold,Chang2,Gezerlis,Lee,Adhikari,Leyronas} and non-zero temperatures \cite{Nozieres,SadeMelo}, still, important issues remain to be discussed. One of them refers to the recently identified {\it Contact} variable \cite{Tan}, which besides from being amenable to direct measurement\cite{Stewart}, yields information on a host of different properties, among them, density-density correlations \cite{Book-BEC-BCS}, the number of pairs in the closed channel of a Feshbach resonance\cite{Castin,Zhang}, and on the forms of thermodynamic variables at unitarity \cite{Tan}. The main issue under scrutiny here is the fact that the calculation of the Contact is very sensitive to different approximations. In particular, it has been argued\cite{Castin,Levinsen} that the usual mean-field approach is inappropriate since it predicts divergent behavior of several thermodynamic variables at unitarity in contrast to experimental evidence\cite{Partridge,Stewart} and more sophisticated approaches\cite{Book-BEC-BCS}. We shall show that mean field remains correct even at unitarity and that the trouble is caused not  by the ansatz itself but by not taking into account the finiteness of the interatomic potential range. To address this point, we shall study 
 the crossover within the BCS mean-field theory (MF) but {\it without} making the usual contact approximation in the interatomic interaction potential.
  Namely, we will calculate the thermodynamics at zero temperature for a realistic finite range potential. Although there have been previous analysis with finite-range interatomic potentials\cite{Meera,Chin}, the issues addressed here have not been discussed. 

The system is a balanced mixture of two monoatomic fermionic species with the same atomic mass $m$, and whose interatomic potentials is $U(|r-r'|)$, namely,
\begin{eqnarray}
H&=&\sum_k \epsilon_k^{\phantom{\dagger}}(\hat a_k^{\dagger}\hat a_k + \hat b_k^{\dagger}\hat b_k^{\phantom{\dagger}})
+ \frac{1}{V}  \sum_{k k' q} \tilde U_q^{\phantom{\dagger}} 
 \hat a_{k+q}^{\dagger}
 \hat b_{k'-q}^{\dagger} 
 \hat b^{\phantom{\dagger}}_{k'}  
 \hat a^{\phantom{\dagger}}_{k}
\nonumber \\
&+&\frac{1}{2V}\sum_{k k' q} \tilde U_q^{\phantom{\dagger}}[
\hat a_{k+q}^{\dagger}
\hat a_{k'-q}^{\dagger}
\hat a^{\phantom{\dagger}}_{k'} 
\hat a^{\phantom{\dagger}}_{k} + 
\hat b_{k+q}^{\dagger}
\hat  b_{k'-q}^{\dagger}
\hat b^{\phantom{\dagger}}_{k'}
\hat b^{\phantom{\dagger}}_{k} ]
  \label{H}
\end{eqnarray}
where $V$ is the volume that confines the sample, $\epsilon_k= \hbar^2k^2/2m$ is the energy of a particle with momentum $\pm \hbar k$. $\hat a_k$ and $\hat b_k$ are fermionic annihilation operators of the two different species. $\tilde U_q$ is the Fourier transform of the interatomic potential $U(r)$. Here, for simplicity we have assumed that the interatomic potential is the same for any pair of fermions.

The mean field approach is based on the BCS ansatz \cite{BCS,Eagles,Leggett-80} for the ground state of the gas, $|\Psi_\mathrm{BCS}\rangle= \prod_k \> ( u_k^{\phantom{\dagger}} +v_k^{\phantom{\dagger}} \hat a_{k} ^\dagger\hat  b_{-k} ^\dagger) \> |0\rangle$
with $u_k$ and $v_k$ variational parameters satisfying $u_k^2+v_k^2=1$ for normalization of the state. A straightforward evaluation of the grand potential $\Omega= \langle \Psi_\mathrm{BCS} | H - \mu [ \sum_k( \hat a_k^\dagger \hat a^{\phantom{\dagger}}_k +\hat b_k^\dagger \hat b^{\phantom{\dagger}}_k)] |\Psi_\mathrm{BCS}\rangle$, where the common chemical potential $\mu$ for both species ensures a balanced mixture, yields
 \begin{equation}
\Omega=2\sum_k ( \epsilon_k^{\phantom{\dagger}}-\mu + \epsilon^{HF}_k) v_k^2
+\frac{1}{V}  \sum_{k k'} \tilde U_{k-k'}^{\phantom{\dagger}} F_k^{\phantom{\dagger}} F_{k'}^{\phantom{\dagger}},  \label{Omega}
\end{equation}
where $F_k=u_k v_k$. The term $\epsilon^{HF}_k$ stands for the Hartree and Fock  energy contributions, given by,
\begin{equation}
\epsilon^{HF}_k=\frac{\tilde U_0}{V}\sum_{k'} v_{k'}^2 - \frac{1}{2V}\sum_{k'} \tilde U_{k-k'}v_{k'}^2 .\label{HF}
\end{equation}
The variational scheme yields the gap equation,
\begin{equation}
\Delta_k=-\frac{1}{V}\sum_{k'} \tilde U_{k-k'} \frac {\Delta_{k'}}{2E_{k'}},
\label{D}
\end{equation}
where the quasiparticle excitation energy is $E_k= [(\epsilon_k-\mu+\epsilon^{HF}_k)^2+\Delta_k^2]^{1/2}$. In addition, the total number of particles is $N = 2 \sum_k v_k^2$. The above equations are valid for any short range interatomic potential.
The contact approximation\cite{Leggett-80} is implemented by setting all the momenta dependence as $\tilde U_q  \approx \tilde U_0 \equiv 4 \pi \hbar^2 a/m$, and as it is well known, such a procedure inevitably leads to convergence difficulties at large values of the momentum $q$, with the necessity of either schemes of renormalization nature or introducing physically justified cut-off parameters\cite{Leggett-80,Gorkov,Bulgac2,Nikolic}. Moreover, as it has been widely indicated in the literature, see for instance Ref.\cite{Levinsen}, the contact approximation only holds in the weakly interacting regime $N |a|^3 /V \ll 1$, and it should therefore not be correct at unitarity where the scattering length diverges. It is certainly surprising that the contact approximation in the BCS ansatz but neglecting Hartree-Fock (HF) terms, which are linearly proportional to $a$, see Eq.(\ref{HF}), gives a finite and smooth crossover at unitarity\cite{Engelbrecht}, not only of the Contact but of the whole thermodynamics. On the other hand, it has been clearly noted\cite{Castin} that inclusion of HF terms indeed shows MF is incorrect since, in particular, gives a divergent Contact at unitarity. All these observations should suffice to disregard MF as a valid and useful theory in the strongly interacting regime. Our simple observation is that the interatomic potential is truly not a linear function of $a$, this is a mathematical device indeed only justifiable in the weakly interacting regime\cite{Levinsen}. In general, what appears in the HF and BCS terms in the grand potential (\ref{Omega}) is the Fourier transform $\tilde U_q$ of the potential. This quantity never diverges, even in the strong interacting regime, if the finite range of the interaction is kept. Accordingly, MF yields always finite results even if HF terms are kept. We shall show this in detail below and, in particular, we will show that in the deep BCS and BEC regimes MF yields expected correct asymptotic results and, at unitarity, reasonable agreement with more sophisticated calculations. 

In order to avoid the contact approximation we use a physically acceptable interatomic potential, an exponential-type of potential,
\begin{equation}
U(r)=-V_0e^{-r/\sigma}
\label{rarita}
\end{equation}
where $\sigma$ is the interaction range and $V_0$ the depth of the potential; its Fourier transform can be analytically calculated.
This potential has essentially two main virtues. First, as a realistic potential it depends on at least two parameters, $V_0$ and $\sigma$. Second, but more important for our purposes, as derived by Rarita\cite{Rarita}, the two-body scattering length is exactly known,
\begin{equation}
a=-2\sigma \left[ \frac{\pi}{2} \frac{N_0(x)}{J_0(x)} - \ln \left(\frac{x}{\sqrt{2}}\right) -\gamma \right],\label{a-R}
\end{equation}
where $x=\sigma \sqrt{2V_0m}/\hbar$, $J_0(x)$ and $N_0(x)$ are zero order Bessel functions of first and second kind and $\gamma$ is the Euler-Mascheroni constant. Therefore, for fixed range $\sigma$, the Grand Potential $\Omega$ depends on the scattering length via $V_0$.  As $V_0$ varies, $a$ passes through an infinite series of scattering resonances where $a$ diverges changing abruptly its sign; the resonances occur as actual bound states emerge\cite{Rarita}. As it has also been widely discussed,  those potential resonances are good approximations to experiments  since essentially all crossover experiments are in the so-called broad Feshbach resonance limit where the width of the resonance is much larger than the Fermi energy\cite{Book-BEC-BCS,Chin,Kohler}. In this study we consider the BCS-BEC crossover through the first resonance. The results of the contact approximation are recovered in the limit $\sigma \to 0$, as shown below.

To analyze the thermodynamics in the whole BCS-BEC crossover, namely, as the {\it inverse} scattering length varies from $-\infty$ to $\infty$, we numerically solve\cite{details} the set of equations (\ref{D})-(\ref{a-R}) without any further assumption or approximation. 
We determine the chemical potential $\mu$, the pressure $p = - \Omega/V$ and the Contact variable ${\cal C}$ given by\cite{Tan,Castin,VRR}
\begin{equation}
{\cal C} = - \left(\frac{\partial \Omega}{\partial \eta}\right)_{\mu, V, T} ,\label{C}
\end{equation}
where $\eta = 1/a$ is the inverse scattering length. Dimensionless variables are formed with Fermi energy and momentum, 
$\epsilon_F=\hbar^2 k_F^2/2m$ and $k_F= \left( 3\pi^2{N}/{V} \right)^{1/3}$, and denoted with a tilde. 

\begin{figure}
\begin{center}
 \includegraphics[width=0.75\textwidth]{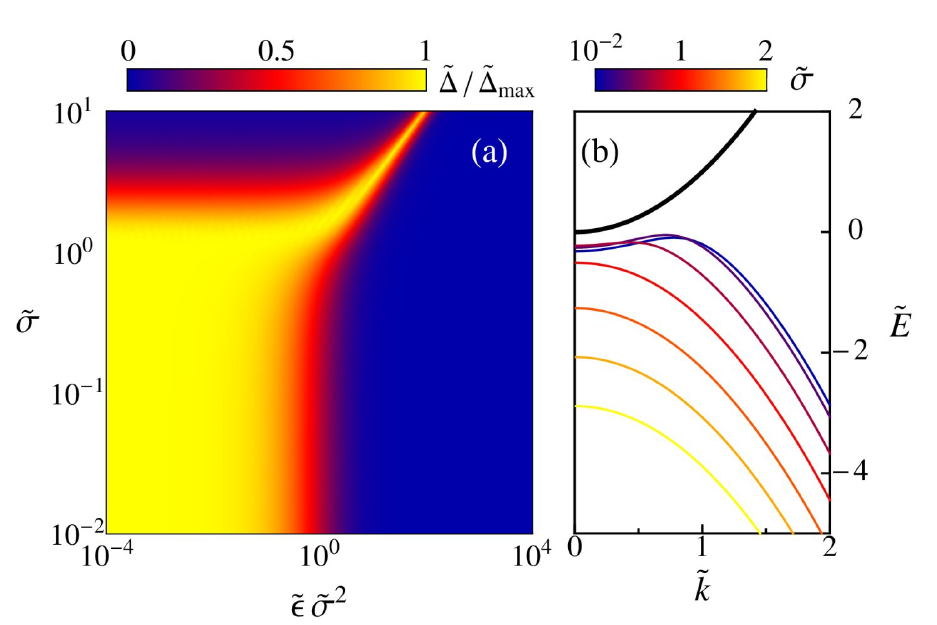}
\end{center}
\caption{(color online)
 (a) Dimensionless gap $\tilde \Delta / \tilde \Delta{\mathrm{max}}$ as a function of interatomic interaction range $\tilde \sigma$ and as a function of the scaled energy $\tilde \epsilon \tilde \sigma^2$. (b) Quasiparticle excitation energy, $\tilde E=\tilde{\mu}-\tilde E_{\tilde k}$, for different values of the potential range $\tilde \sigma$. The thick line corresponds to $\tilde k^2$. These calculations correspond to $\tilde \eta =0$.}
\label{gap}
\end{figure}

To begin our discussion, we show in Fig.\ref{gap}(a) the energy dependent gap, Eq.(\ref{D}), $\tilde \Delta(\tilde \epsilon, \tilde \sigma)/ \tilde \Delta_{\mathrm{max}}(\tilde \sigma)$, normalized to its maximum value.
We show only the strongly interacting regime ($\tilde \eta =0$) but the results are very similar for arbitrary values of the scattering length. We find that as  long as $\tilde \sigma \lesssim 1$, the gap is a constant $\tilde \Delta \approx \tilde \Delta_{\mathrm{max}}$ for $\tilde \epsilon \tilde \sigma^2 \lesssim 1$, then abruptly drops to zero. As $\tilde \sigma \to 0$, we recover a constant gap for all $\tilde \epsilon$. From moderate to large values of $\tilde \sigma$ the gap shows a peak. As we see below, this novel dependence of the gap for large values of $\tilde k$ dramatically changes the behavior of the thermodynamic properties. 

\begin{figure}
\begin{center}
 \includegraphics[width=0.75\textwidth]{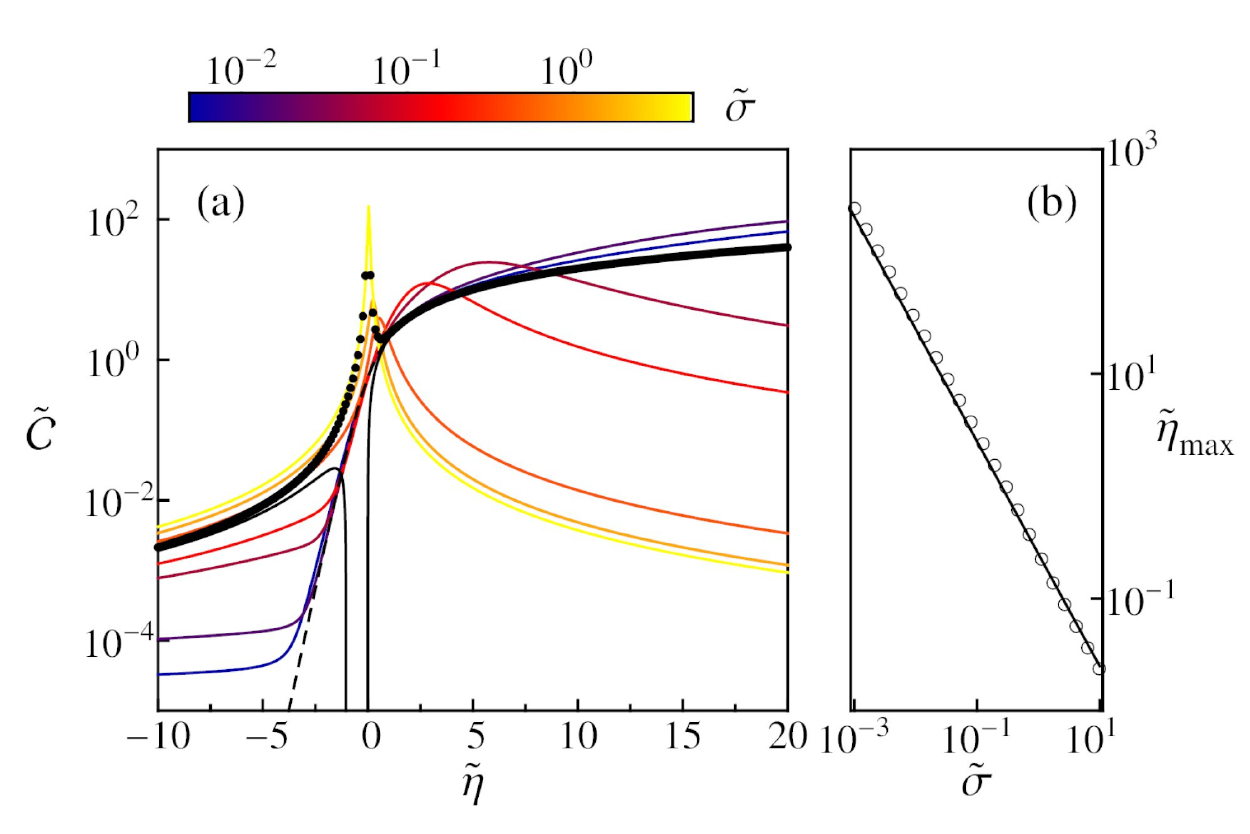}
\end{center}
\caption{(color online)
(a) Dimensionless contact variable $\tilde {\cal C}={\cal C} k_F/(N \epsilon_F)$ as a function of inverse scattering length $\tilde \eta = 1/(k_F a)$. The color curves are the results of the full BCS equations for finite values of $\tilde \sigma$, coded in the colored bar above. The dotted line is the BCS solution in the contact approximation neglecting HF terms. The continuous black lines are HYL limit. (b)  The value of $\tilde \eta_{\mathrm{max}}$ where the contact takes its maximum value.  It obeys $\tilde \eta_{\mathrm{max}} \approx 1/(4 \tilde \sigma)$.}
\label{contact}
\end{figure}

We now  address the main point of this letter. This is the behavior of the contact variable ${\cal C}$, eq. (\ref {C}), for the whole crossover at $T=0$.  
Fig. \ref{contact}(a) shows $\tilde {\cal C}={\cal C}k_F/(N\epsilon_F)$ as a function of $\tilde \eta = 1/(k_F a)$ for different values of the range $\tilde \sigma$. First, we highlight the dashed black line that corresponds to the solution of the BCS equations in the contact approximation but completely {\it neglecting} the Hartree-Fock (HF) terms, see eq. (\ref{Omega}). This approximation has been deemed as inappropriate\cite{Castin} to describe the system since in the deep BCS regime, $\tilde \eta \to -\infty$, it predicts an exponential decay, as can be seen in Fig. \ref{contact}(a), yet it gives a smooth finite crossover at unitarity. The correct behavior of $\tilde {\cal C}$ is believed to be given by the weakly-interacting approximations of a Fermi liquid in the BCS side and by the diatomic version of a Bose gas with repulsive interactions in the BEC side, these in turn, given by the celebrated expressions of Huang-Yang-Lee (HYL)\cite{HYL}. In the very deep BCS and BEC regimes, HYL  yields as leading terms $\tilde {\cal C} \approx 2/(3 \pi\tilde \eta^{2})$ in the BCS side, and  $\tilde {\cal C}\approx 2 \tilde \eta$ in the BEC extreme. These expressions are shown with solid black lines in Fig. \ref{contact}(a) becoming, the former zero and the latter divergent at unitarity. In addition to these lines, with black dots, we also plot the mean-field BCS theory in the contact approximation {\it with} HF terms. We see that as $\tilde \eta \to \pm \infty$, this solution approaches HYL result but, at the crossover $\tilde \eta = 0$, it also diverges. This is one of the purported indications that MF is invalid at the crossover. As we now show, these divergences are due to the contact approximation and not to MF approach itself. In color lines we show the finite range calculations. 
In general, we find that if the interaction range is small, $\tilde \sigma \lesssim 1$, the contact $\tilde {\cal C}$ eventually vanishes algebraically as $ \tilde \eta^{-2}$ in the BCS side ($\tilde \eta < 0$) in agreement with HYL. However, very interestingly, one can see that as $\tilde \sigma$ decreases, such a behavior is delayed for larger negative values of $\tilde \eta$, indicating that in the limit $\tilde \sigma \to 0$ the finite range calculation approaches that predicted by BCS theory {\it with} contact approximation but {\it without} HF terms.
 
 In the BEC side a novel behavior emerges due to the finite character of  $\tilde \sigma$.  The contact variable $\tilde {\cal C}$ peaks at a value that depends on $\tilde \sigma$ and then decays to zero again as $\tilde \eta$ increases. The value of $\tilde \eta_{\mathrm{max}}$ at the peak scales simply as $\tilde \sigma^{-1}$, see Fig. \ref{contact}(b). We further note that as $\tilde \sigma$ keeps increasing, the location of the peak not only tends to $\tilde \eta = 0$, i.e. to unitarity, but the value of $\tilde {\cal C}$ becomes sharply peaked, suggesting that the crossover at unitarity becomes a true quantum phase transition. Although the transition may appear unrealistic since one needs an infinite interatomic range, we would like to recall that the present theory is a mean-field one. Such a transition is reminiscent of the Kac limit\cite{Kac} of the van der Waals fluid. 

\begin{figure}
\begin{center}
 \includegraphics[width=0.75\textwidth]{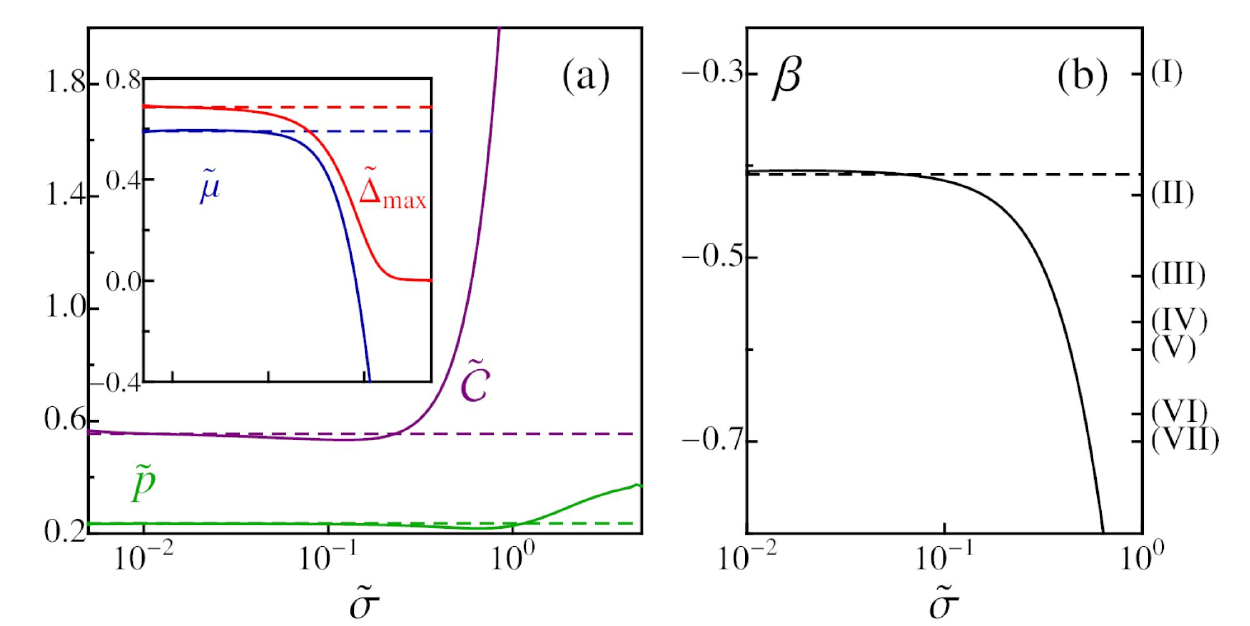}
\end{center}
\caption{(color online)
Thermodynamics at unitarity $\tilde \eta = 0$. (a) The chemical potential $\tilde \mu$, the maximum value of the gap $\tilde \Delta_{\mathrm{max}}$, the pressure $\tilde p$ and the contact variable $\tilde {\cal C}$, as functions of the interaction range $\tilde \sigma$.  (b) Bertsch parameter $\beta$ as a function of $\tilde \sigma$. The numbers on the right correspond to references of different calculations, 
(I)$\to$\cite{Bruun}, 
(II)$\to$\cite{Bertsch}, 
(III)$\to$\cite{Bruun,Nishida},
(IV)$\to$\cite{Carlson,Chang,Carlson2, Astrakharchick,Chang2, Bulgac2},
(V)$\to$\cite{Hu,Arnold}, 
(VI)$\to$\cite{Lee,Heiselberg,Bertsch}, and 
(VII)$\to$\cite{Arnold,Lee}  . Dashed lines correspond to the predicted values at unitarity of the BCS theory in the contact approximation {\it without} the Hartree-Fock terms.}
\label{unitarity}
\end{figure} 

A very important issue in this discussion is certainly the unitarity region where the scattering length diverges $a \to \pm \infty$, signaling the appearance of a resonance and giving rise to very strong interatomic interactions.
As mentioned above, if one sticks to the strict contact approximation, certainly the model breaks down since the Hartree-Fock terms become divergent. However, for finite values of $\tilde \sigma$, HF terms are perfectly finite even if $a$ becomes unbounded. Since our calculation never really uses the contact approximation one never faces any divergence. The resulting crossover is smooth as one would expect, except when $\tilde \sigma$ grows indefinitely. Fig. \ref{unitarity} shows the thermodynamics at unitarity. First, we note that for $\tilde \sigma \lesssim 1$, the exact calculation approaches that of the usual BCS mean-field calculation, {\it with} the contact approximation but {\it without} including the divergent Hartree-Fock terms, dotted lines in the figure. That is, for small interatomic range, all thermodynamic quantities approach MF universality class\cite{Ho}. For larger values of $\tilde \sigma$ evident deviations appear. In the figure we also show Bertsch parameter $\beta$\cite{Bertsch} for the present calculation together with a collection of corresponding values for different type of models and calculations\cite{Heiselberg,Carlson,Chang,Carlson2,Bruun,Astrakharchick,Hu,Nishida,Arnold,Chang2,Lee,Bertsch}. We clearly see that all those values are within $\tilde \sigma \lesssim 1$ of the exact calculation. Full elucidation of this important point is beyond the scope of this article. 

This work was partially supported by grant IN108812-2 DGAPA (UNAM).

\end{document}